\documentclass[twocolumn,twoside,slac_two]{revtex4}
\usepackage{graphicx}
\usepackage{amsmath}
\usepackage{fancyhdr}
\begin{document}

\title{A Combined Analysis of Clusters of Galaxies - Gamma-Ray Emission from Cosmic Rays and Dark Matter}

%

\author{S. Zimmer, J. Conrad}
\affiliation{The Oskar Klein Centre for Cosmoparticle Physics, AlbaNova, SE-10691 Stockholm, Sweden and Department of Physics, Stockholm University, AlbaNova, SE-10691 Stockholm, Sweden}
\author{for the \textit{Fermi}-LAT Collaboration}
\noaffiliation
\author{\bigskip A. Pinzke}
\affiliation{Department of Physics, Broida Hall, University of California Santa Barbara, CA 93106-9530, USA}

\begin{abstract}
Multiwavelength observations suggest that clusters are reservoirs of vast amounts of relativistic electrons and positrons that are either injected into and accelerated directly in the intra-cluster medium, or produced as secondary pairs by cosmic ray ions scattering on ambient protons. In these possible scenarios gamma rays are produced either through electrons upscattering low-energy photons or by decay of neutral pions produced by hadronic interactions. In addition, the high mass-to-light ratios in clusters in combination with considerable Dark Matter (DM) overdensities makes them interesting targets for indirect DM searches with gamma rays.\\
The resulting signals are different from known point sources or from diffuse emission and could possibly be detected with the \textit{Fermi}-LAT. Both WIMP annihilation/decay spectra and cosmic ray induced emission are determined by universal parameters, which make a combined statistical likelihood analysis feasible.
We present initial results of this analysis leading to limits on the DM annihilation cross section or
decay time and on the hadron injection efficiency.
\end{abstract}

\maketitle

\thispagestyle{fancy}

\section{Introduction}
Gamma rays from clusters of galaxies have not yet been detected\footnote{However, galaxies who are members of clusters have been detected.}. On the other hand, from multiwavelength observations it is known that clusters have a very rich astrophysical phenomenology and observations support the assumption of relativistic particle populations in the intra-cluster medium (ICM). These particle populations can interact and produce high energy gamma rays \cite{Pinzke2011}, potentially being detectable  by the Large Area Telescope (LAT) on board the \textit{Fermi} satellite \cite{Atwood2009}. 

Present day observations \cite{Amanullah2010,Jarosik2011} suggest the content of the universe to be made of about 25\% of Dark Matter (DM), whose nature however remains unknown. DM annihilation or decay into standard model particles can lead to a gamma-ray flux, detectable by \textit{Fermi}-LAT \cite{Baltz2008}. The annihilation flux is proportional to the density squared, therefore galaxy clusters with substructure which can lead to large enhancements of the DM induced gamma-ray flux are promising targets for Dark Matter seaches.

Both CR and DM-induced emission scenarios are different in spectral shape from any of the observed signals measured by the LAT, such as the emission from identified point sources or the galactic and extragalactic diffuse emission. Assuming both the cosmic ray (CR) and DM emission model being universal in at least one defining parameter in all clusters we study, we can use a combined likelihood analysis to constrain the parameters that define those emission models.

Both CR and DM induced gamma-ray emission need to be considered together in an attempt to quantify a possible DM or CR detection. In this paper, we present preliminary results where both components are considered separately.

The outline of this proceedings is as follows. In Section 2 we discuss the cluster selection and their modeling with respect to the two emission models given by DM and CR, respectively. We present the analysis details and our combined likelihood method in Section 3. The results from the upper limit evaluations are described in Section 4. 

\section{Cluster Selection and Modeling}
The clusters were selected based on the same criteria as in the first \textit{Fermi}-LAT publication on DM induced gamma rays in galaxy clusters \cite{Ackermann2010}. We chose the brightest X-ray clusters from the extended HIFLUCS catalog \cite{Reiprich2002} and removed clusters that were near the galactic plane (Norma and Ophiuchus) as well as clusters with detected active galaxies in gamma rays, such as Perseus with its central galaxy NGC 1275. In addition we removed NGC 4636 since it exhibits a significant overlap in the region of interest with M49 and our method is at present not capable of treating that problem properly.

It is worth mentioning that due to the lack of gamma-ray emission from any of these objects, we refer to the position in the sky which corresponds to the NASA/IPAC Extragalactic Database locations of clusters as seen in other wavebands.

\subsection{Gamma-Ray Emission from Cosmic Rays}
\begin{figure}
  \includegraphics[width=0.45\textwidth]{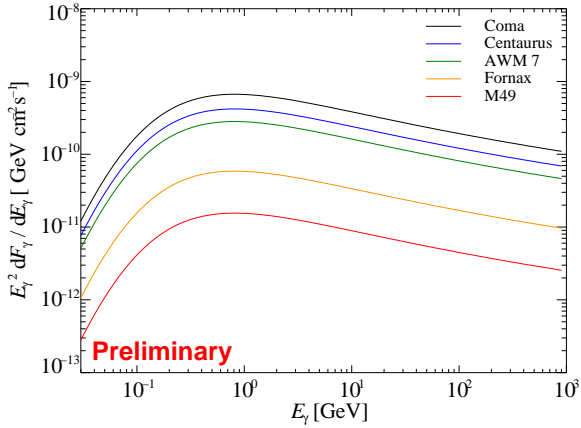}
  \caption{\label{fig:gamma_pinzke} Predicted gamma-ray emission from neutral pion decay as in \cite{Pinzke2010}. We note that Coma is predicted to give the strongest signal.}
\end{figure}

Although there is no clear observational evidence yet for a relativistic proton population in clusters of galaxies, these objects are expected to contain significant populations of CR protons originating from different sources, such as structure formation shocks, radio galaxies, and supernovae driven galactic winds (see e.g. \cite{Volk1996,Ensslin1997,Pfrommer2004,Pinzke2010}). The cluster gamma-ray emission is crucial in this respect as it potentially could provide unique and unambiguous evidence of a CR proton population in clusters through observing the pion bump in the gamma-ray spectrum. High resolution simulations of galaxy clusters including CR physics show that these processes can be described by an approximately universal spectrum, where the gamma-ray flux from CR proton induced neutral pions decaying can be parametrized as:
\begin{equation}
\Phi_{\gamma}^{CR} = \int \mathrm{d}^3r A(R) \lambda_{\pi^{0}\rightarrow \gamma}\left(E\right)
\end{equation}
where $\lambda_{\pi^{0}\rightarrow \gamma}\left(E\right)$ is the universal spectrum as given in \cite{Pinzke2010} containing the maximum hadron injection efficiency, $\eta$, while $A(R)$ is the individual normalization that varies with respect to the mass distribution as inferred e.g. from X-ray observations and its gas density profile \cite{Cote2003}. In fig.~\ref{fig:gamma_pinzke} we show the predicted gamma-ray emission from neutral pion decay for the clusters we chose to investigate.

\subsection{Gamma-Ray Emission from Dark Matter}
The gamma-ray flux from a pair-annhihilating DM particle with mass $m_{\mathrm{WIMP}}$ can be written as:
\begin{equation}\label{Eq:DM_A}
  \Phi_{\gamma}^{A} = \frac{1}{2} \frac{\langle \sigma v \rangle}{m_{\mathrm{WIMP}}^2} \sum_{f}{\frac{\mathrm{d}N_{f}}{\mathrm{dE}}\cdot B_{f}} \times\overline{J}_{A},
\end{equation}
where $\langle \sigma v \rangle$ is the DM annhihilation cross section, and $\sum_{f}{\frac{\mathrm{d}N_{f}}{\mathrm{dE}}\cdot B_{f}}$ denotes the gamma-ray yield for the given final state with branching ratio $B_f$. This particle physics factor is multiplied with the astrophysical factor, $\overline{J}_{A}$ that in case of annhihilating DM is the integral over the line of sight and the DM density squared over a specified solid angle. For decaying DM we use the fact that the spectrum is roughly equivalent to the annhihilation spectrum of a particle with half the mass. Furthermore we note that the astrophysical factor, $\overline{J}_{D}$ is only integrated over the DM density, not the squared one, in the case of decay. Eq.~\ref{Eq:DM_A} then becomes:
\begin{equation}\label{Eq:DM_D}
  \Phi_{\gamma}^{D} = \frac{1}{m_{\mathrm{WIMP}}\tau} \sum_{f}{\frac{\mathrm{d}N_{f}}{\mathrm{dE}}\cdot B_{f}} \times\overline{J}_{D},
\end{equation}
where we denote the lifetime of the DM particle with $\tau$.

\section{Analysis Details}
We carry out a binned analysis of 24 months of \textit{Fermi}-LAT data. We select events corresponding to the {\tt DIFFUSE} class which are most gamma-ray--like events and use the {\tt P6\_V11} instrument response function, provided in the software package {\tt Fermi Science Tools v9r22p0}. The spectra are binned in 20 logarithmic bins from 200~MeV to 100~GeV. 

According to the model predictions for CR-induced gamma rays from galaxy clusters, the bulk of the emission comes from the center of the cluster which justifies our choice of modeling them as pointlike sources. For the case of DM we assume the branching fraction to be 100~\% to $\mathrm{b\overline{b}}$ both for decay and annhihilation, which is shown to be a good approximation for a large class of possible DM models \cite{Cesarini2002,Morselli2002}. When modeling the clusters we model them as point sources using the same J-factors as in \cite{Ackermann2010} for annhihilating DM and \cite{Dugger2010} for decaying DM, without assuming any boost from substructure.

We include all known point sources from the first year \textit{Fermi}-LAT catalog \cite{Abdo2010} that are within 15 degrees from the location of each cluster using \textit{Fermi}-LAT catalog fit results,  except for point sources within 5 degrees from the source, which are refitted. In addition we model the diffuse and extragalactic gamma-ray background according to the recommended \textit{Fermi}-LAT templates\footnote{For the galactic diffuse emission we use the template {\tt gll\_iem\_v02\_P6\_V11\_DIFFUSE.fit} and for the isotropic gamma-ray background {\tt isotropic\_iem\_v02\_P6\_V11\_DIFFUSE.txt} respectively.}.

When we inspect the spectra and the residuals of individual sources with respect to the assumed model, we conclude that the measured $\gamma$-ray emission is modeled well by assuming only detected \textit{Fermi}-LAT point sources and diffuse emission provided by the galactic and extragalactic diffuse models respectively.

\subsection{Likelihood Procedure}
In order to model each of the cluster regions given its own characteristics together with the underlying universal physics as assumed for emission from DM or CR respectively, we construct a combined likelihood function by multipylying every individual likelihood function and minimizing it with respect to the parameter of interest, e.g. the annhihilation cross section, the decay time or the hadron injection efficiency. The likelihood function can then be written as:
\begin{equation}\label{Eq:compLike}
\begin{split}
L&\left(\langle \sigma v\rangle,m_{\mathrm{WIMP}}|obs\right) \\
&=\prod_{i} L_{i}\left(\langle \sigma v\rangle,m_{\mathrm{WIMP}},c_{i},b_{i}|obs\right).
\end{split}
\end{equation} 
In Eq.~\ref{Eq:compLike} we use for simplicity the case of annhilating DM. For a given mass of a WIMP we can minimize this function and extract upper limits on the annhihilation cross section $\langle \sigma v\rangle$. The $b_{i}$ correspond to individual backgrounds that are treated as nuisance parameters in the likelihood evaluation, the $c_i$ represent individual constants such as cluster masses. 

This method is implemented in the {\tt Fermi Science Tools} as {\tt Composite2} and profiling over the likelihood function is achieved by means of MINOS which is part of the MINUIT package \cite{James1975}. It should be stressed that we refer to the 95~\% upper and lower limits as the ones that are defined as the points where the negative logarithm of likelihood function has changed its value by 2.71 with respect to its minimum. The method allows inclusion of the J factor uncertainties, see also \cite{Ackermann2011} for more details. In order to do so, the distributions of the J-factors need to be known. However, since the distribution of J-factors depends on underlying assumptions such as whether or not DM substructure is included and the details of its assumed distribution and properties, we refrain from including J-factor uncertainties in our analysis. 

\section{Results}
\begin{figure}
  \includegraphics[width=0.5\textwidth]{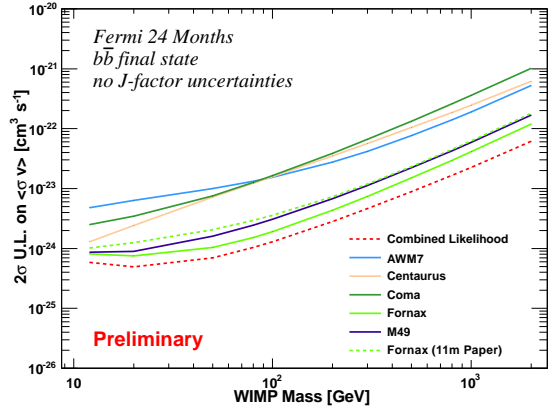}
  \caption{\label{fig:sigmav}95~\% upper limits on the Dark Matter annhihilation cross section for individual clusters (solid lines) and combined result (dotted) for $b\overline{b}$ final states. For comparison we also show the published values for the best cluster in the 11~months analysis (dashed) \cite{Ackermann2010}.}
\end{figure}
\begin{figure}
  \includegraphics[width=0.5\textwidth]{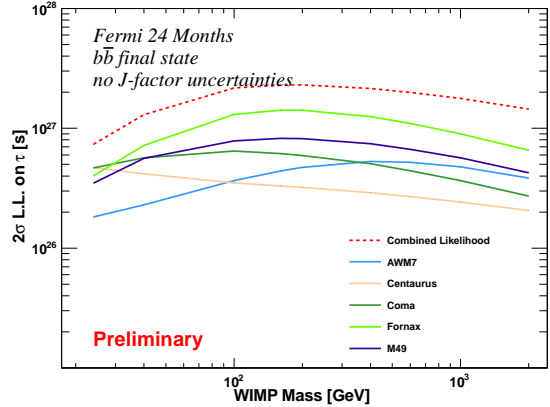}
  \caption{\label{fig:tau}95~\% lower limits on Dark Matter decay time. Same as fig~\ref{fig:sigmav}.}
\end{figure}
\begin{figure}
  \includegraphics[width=0.5\textwidth]{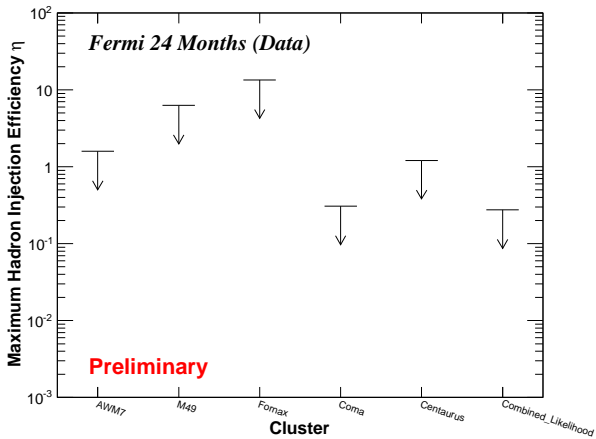}
  \caption{\label{fig:eta}95~\% upper limits on maximum injection efficiency following the hadronic cosmic ray model by \cite{Pinzke2010}. The rightmost point indicates the combined constraint.}
\end{figure}

We show the upper limits on the annhihilation cross section for our cluster sample and the combined result in fig.~\ref{fig:sigmav}. Below 100 GeV we exclude gamma rays from annhihilating DM with a annhihilation cross section above $5\times10^{-25}\mathrm{cm}^{3}\mathrm{s}^{-1}$, while we note that our combined limit (dashed red line) is roughly a factor 3 more constraining than the best individual cluster (Fornax). For the case of decaying DM we exclude life times below $10^{27}~\mathrm{s}$ for WIMPs decaying into $\mathrm{b\overline{b}}$ final states with masses above 100~GeV, as we show in fig~\ref{fig:tau}.
The adopted CR model assumes the maximum injection efficiency $\eta$ to be 50 \%. We find that while only the case of Coma can constrain this parameter space, since the limit we find is below 50~\%, that our data disfavors this value, see fig.~\ref{fig:eta}. Assuming a linear relation between the CR-induced gamma-ray flux and $\eta$, we can establish a combined upper limit on $\eta$ of 27~\%, given the model characteristics by \cite{Pinzke2010}. 

\section{Summary and Conclusions}
Clusters of galaxies provide very interesting targets for searches for gamma-ray emission from both hadronic CR interactions in the ICM and from Dark Matter annhihilation or decay. However, there is no confirmed observation of gamma rays from clusters so far. 

Individual fits of \textit{Fermi}-LAT data in the region of 5 clusters is compatible with identified \textit{Fermi}-LAT point sources and galactic and extragalactic diffuse emission. From the non-detection of clusters in gamma rays, we can obtain upper limits on Dark Matter annhihilation cross section or decay time for a generic WIMP with a $b\overline{b}$ final state. For the discussed emission model from hadronic cosmic ray interactions we are able to set upper limits on the hadron injection efficiency.
 
To capture each region and its individual backgrounds we employ a combined likelihood analysis treating those individual backgrounds as nuisance parameters. This approach is feasible as we assume all clusters to exhibit the same physical properties, which is undisputed for the DM case and can be assumed for the cosmic-ray emission to first approximation. Our combined limits on DM are a factor two better than previously published limits and our data disfavor CR hadrons being accelerated with an efficiency greater than 30 \%. We note that one cannot lower the acceleration efficiency infinitely if one wants to explain radio halos with the hadronic CR model.

It is however noteworthy that this analysis is only a proof-of-concept study to show the power of the combined likelihood analysis. We are currently updating the analysis to increase the number of observed clusters, including more accurate spatial modeling, improved treatement of overlapping ROIs and exploration of other emission scenarios. We are also using an updated \textit{Fermi}-LAT event selection. A corresponding \textit{Fermi}-LAT publication is in preparation.







\bigskip 
\begin{acknowledgments}
The \textit{Fermi} LAT Collaboration acknowledges generous ongoing support
from a number of agencies and institutes that have supported both the
development and the operation of the LAT as well as scientific data analysis.
These include the National Aeronautics and Space Administration and the
Department of Energy in the United States, the Commissariat \`a l'Energie Atomique
and the Centre National de la Recherche Scientifique / Institut National de Physique
Nucl\'eaire et de Physique des Particules in France, the Agenzia Spaziale Italiana
and the Istituto Nazionale di Fisica Nucleare in Italy, the Ministry of Education,
Culture, Sports, Science and Technology (MEXT), High Energy Accelerator Research
Organization (KEK) and Japan Aerospace Exploration Agency (JAXA) in Japan, and
the K.~A.~Wallenberg Foundation, the Swedish Research Council and the
Swedish National Space Board in Sweden.

Additional support for science analysis during the operations phase is gratefully
acknowledged from the Istituto Nazionale di Astrofisica in Italy and the Centre National d'\'Etudes Spatiales in France.
\end{acknowledgments}

\bigskip 
\bibliography{biblio}

\end{document}